\documentclass[12pt,a4paper,dvips]{article}
\usepackage{graphicx}

\newcommand{\Zo}{\mathrm{Z}}
\newcommand{\Zp}{\mathrm{Z'}}
\newcommand{\bq}{\begin{equation}}
\newcommand{\eq}{\end{equation}}

\begin{document}
\begin{center}
{\huge
        Prospects to detect a Z$'$ with the LC
\vspace*{1.cm}
}
\\
{\large
Sabine Riemann}
\vspace*{0.5cm}

\begin{normalsize}
{Deutsches Elektronen-Synchrotron DESY, Institut f\"ur Hochenergiephysik
\\
IfH Zeuthen, Platanenallee 6, D-15738 Zeuthen, Germany }
\end{normalsize}
\end{center}

\vspace*{2.cm}

\begin{abstract} 
\noindent
The search for a Z$'$ is one of the tasks of
future colliders.
I study the capability 
to detect the Z$'$ at a linear e$^+$e$^-$ collider
operating below 
resonance production. 
Depending on $m_{\Zp}$ 
and the collider parameters 
we will be able to discriminate  between Z$'$ models.
\end{abstract}

\section{Introduction}

Despite the excellent agreement of Standard Model predictions 
with present experimental results 
many of us have no doubt that
a more fundamental theory (GUT) describes all forces at high
energy scales by only one gauge group.
The symmetry breaking of a unifying gauge group 
may lead to new gauge bosons at a scale of order 1 TeV.
Popular additional gauge bosons are the Z$'$ coming from an E$_6$ GUT
or the Z$'$=Z$_R$ with W$^{\pm}_R$ arising from symmetry
breaking in
left-right models:
\bq
J_{\Zp}^{\mu} = J_{\chi}^{\mu} \cos \Theta_6 +
                J_{\psi}^{\mu} \sin \Theta_6~;~~~~~
J_{\Zp}^{\mu} = \alpha_{LR} J_{3R}^{\mu}-\frac{1}{2 \alpha_{LR}}
J_{B-L}^{\mu}.
\label{neut_curr}
\eq
Specific cases are the $\chi,~\psi$ and $\eta$ models
($\Theta_6=0; \pi/2;-\arctan \sqrt{5/3}$) and the left-right-models
($\sqrt{2/3}\leq \alpha_{LR} \leq \sqrt{\cot^2 \theta_W-1}$).

With e$^+$e$^-$ colliders the properties of a Z$'$ can be
investigated.
The determination of Z$'$ parameters is easy
if the centre--of--mass energy of a collider
is large enough to produce this  boson.
But even indirect measurements of $e^+e^- \rightarrow
(\gamma, \Zo,\Zp)  \rightarrow \mathrm{f} \bar{\mathrm{f}}$
below the Z$'$ production threshold
gives information about the nature of the Z$'$.

Here, prospective  measurements  of Z$'$  parameters
are presented.
Besides the determination of
the Z$'$ mass
the identification
of the Z$'$ model is reviewed
analyzing fermion-pair production
below a Z$'$ resonance.

I study the following collider scenario:
\bq
\sqrt{s} = ~500 \mathrm{GeV};~~~~~L_{int} = 50 {fb}^{-1} \label{coll}
\eq
The electron beam is polarized, $P_{e^-}=80$~\%.
The results may be extrapolated to the scenarios  $\sqrt{s} = ~800$~GeV,
$L_{int} = 200 $fb$^{-1}$ and $\sqrt{s} = 1600$~GeV,
 $L_{int} = 800 $fb$^{-1}$.

The total cross section, $\sigma_T$, the forward-backward asymmetry,
$A_{FB}$, the left-right asymmetry, $A_{LR}$, and the
forward-backward polarization asymmetry, $A_{LR}^{FB}$, can be measured
with small statistical uncertainties.
For
leptonic and hadronic final states
without discriminating between quark flavors
small systematic errors are expected
(see also \cite{sarisalka}).
The experience of SLD and LEP experiments
has shown
that good techniques of quark flavour identification
with high efficencies and purities are feasible.
However,
background reactions and misidentification of  final state fermions
can lead to relatively large systematic errors
if  q$\bar{\mathrm{ q}}$ final states are analysed.
Hence in the following, the influence of systematic errors
on the determination of Z$'$ parameters is considered.

An angular acceptance cut of 20$^{\circ}$ to final state fermions
is assumed.
Further, the  t-channel exchange in Bhabha scattering
is neglected.
By applying a  cut on the energy
of radiative photons, $\Delta = E_{\gamma}/E_{beam}$,
the radiative return to the Z peak can be suppressed. Here,
$\Delta = 0.9$ is used.
An uncertainty of 0.5\% is taken into account
for the luminosity measurement.
For numerical studies  I use
the program package ZEFIT/ZFITTER \cite{zefit,zfitter}.

\section{Determination of Z$'$ Parameters}

\subsection{Z$'$ Mass}

A crucial element of a Z$'$ search is the determination of its
Z$'$ mass, $m_{\Zp}$. Exclusion limits for $m_{\Zp}$
have been determined for different collider types and various
Z$'$ models. An overview can be found in \cite{zpmass}.

Assuming the collider scenario (\ref{coll}), it is expected that
an analysis of all leptonic and hadronic observables
is sensitive to the  Z$'$ mass as shown in Figure 1 
(see also \cite{alsr,zerwas}).
\begin{figure}
\begin{tabular}{ll}
   \resizebox{!}{7. cm}{%
   \includegraphics{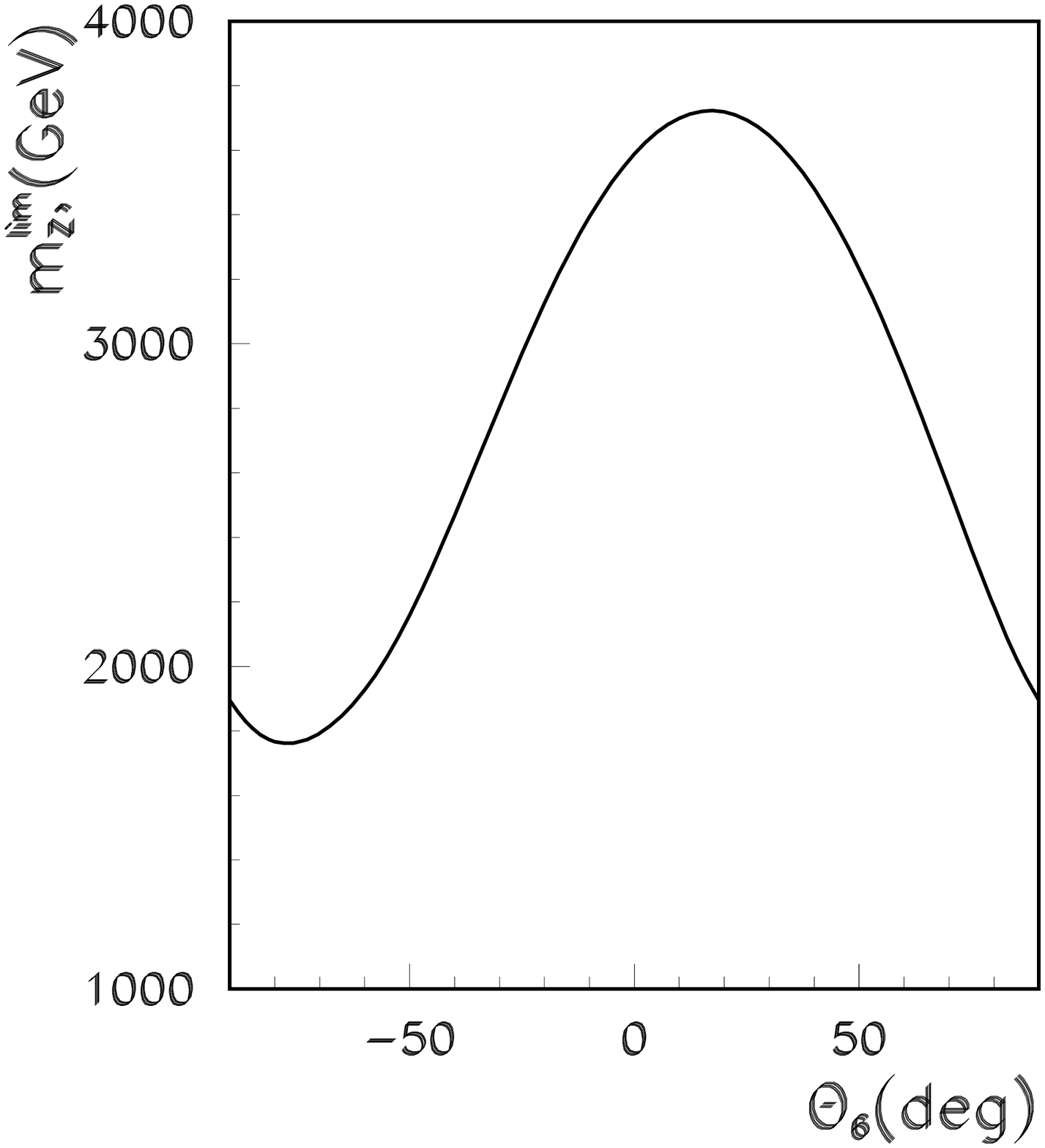}
 }
   &
%
   \resizebox{!}{7. cm}{%
   \includegraphics{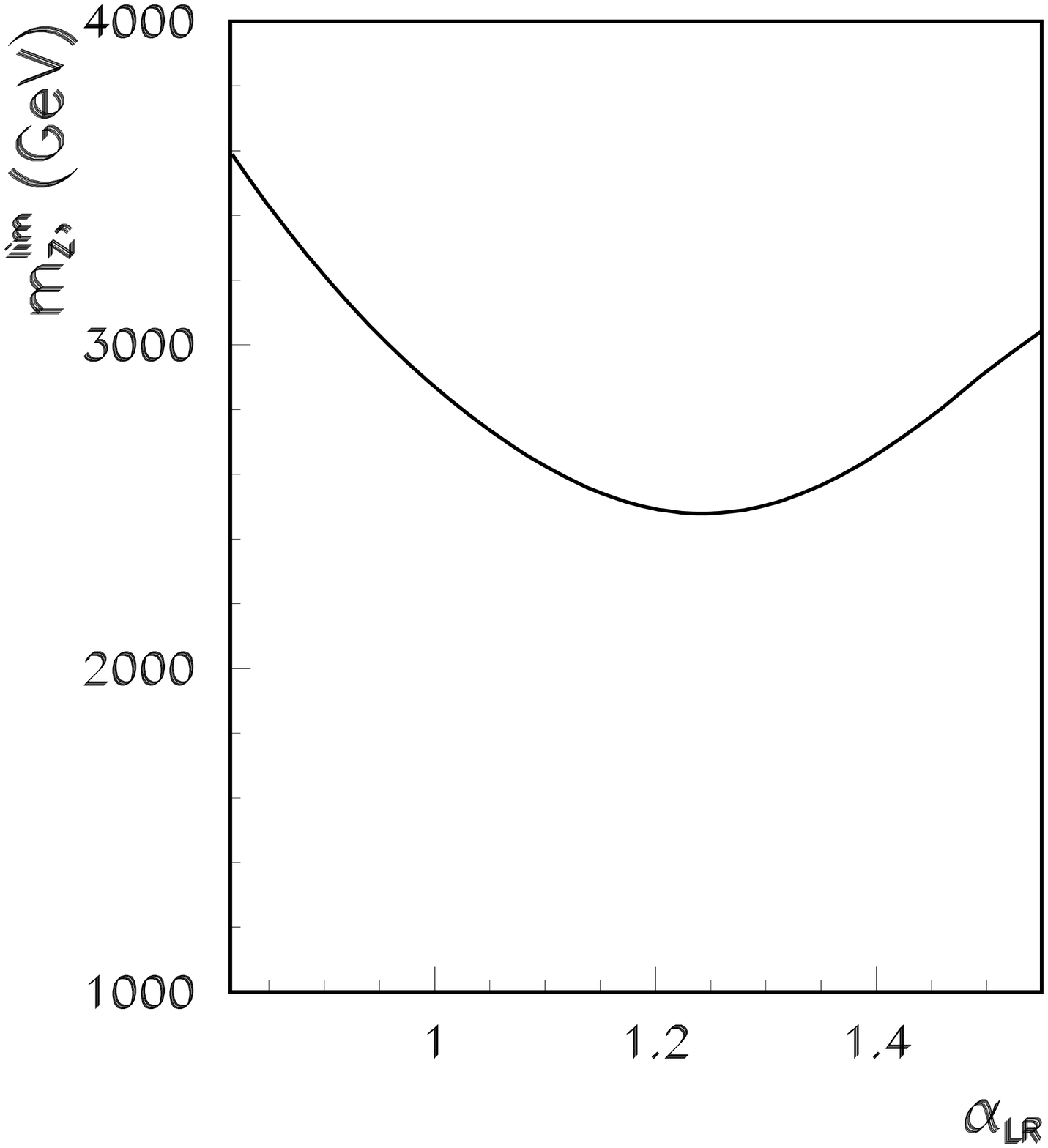}
 }
\end{tabular}
\caption{Lower bounds on the $Z'$ mass, $m_{\Zp}^{lim}$, (95\% CL)
for E$_6$  and left-right models.
}
\label{fig:1}
\end{figure}
here, the Z$'$ model is assumed to be known.

Let us try to determine the Z$'$ mass
assuming the realization of an
E$_6$ GUT
but without any information about the model
parameter $\Theta_6$ in (\ref{neut_curr}).
Figure 2 demonstrates the possibility to
measure $m_{\Zp}$ and $\Theta_6$ simultaneously
if a Z$'$ in the $\chi$ model has a
a  mass of 1 TeV or 1.5 TeV.
%
\begin{figure}

\begin{flushleft}
\begin{tabular}{ll}
\hspace{-1cm}

   \resizebox{!}{7. cm}{%
   \includegraphics{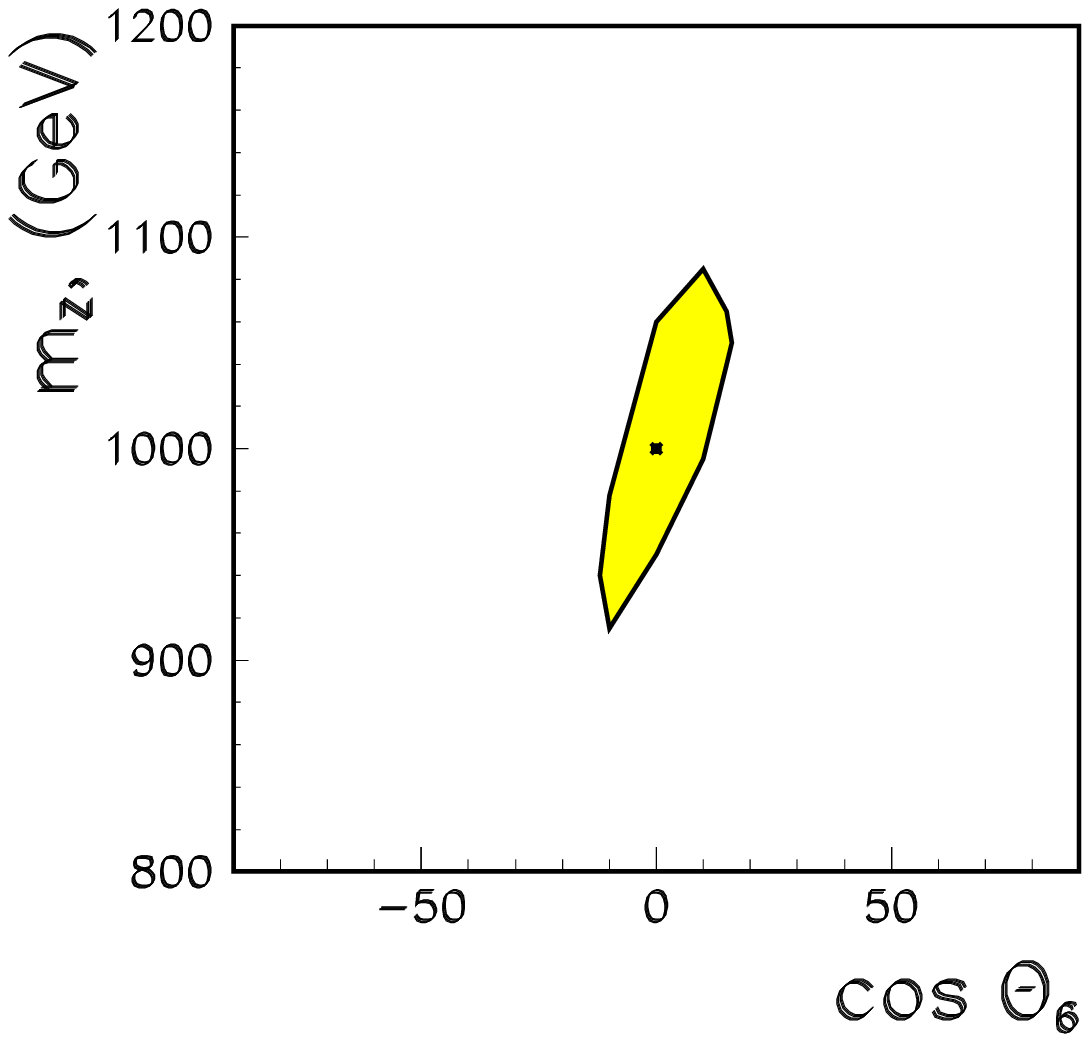}
 }
   &
\hspace{-.51cm}

   \resizebox{!}{7. cm}{%
   \includegraphics{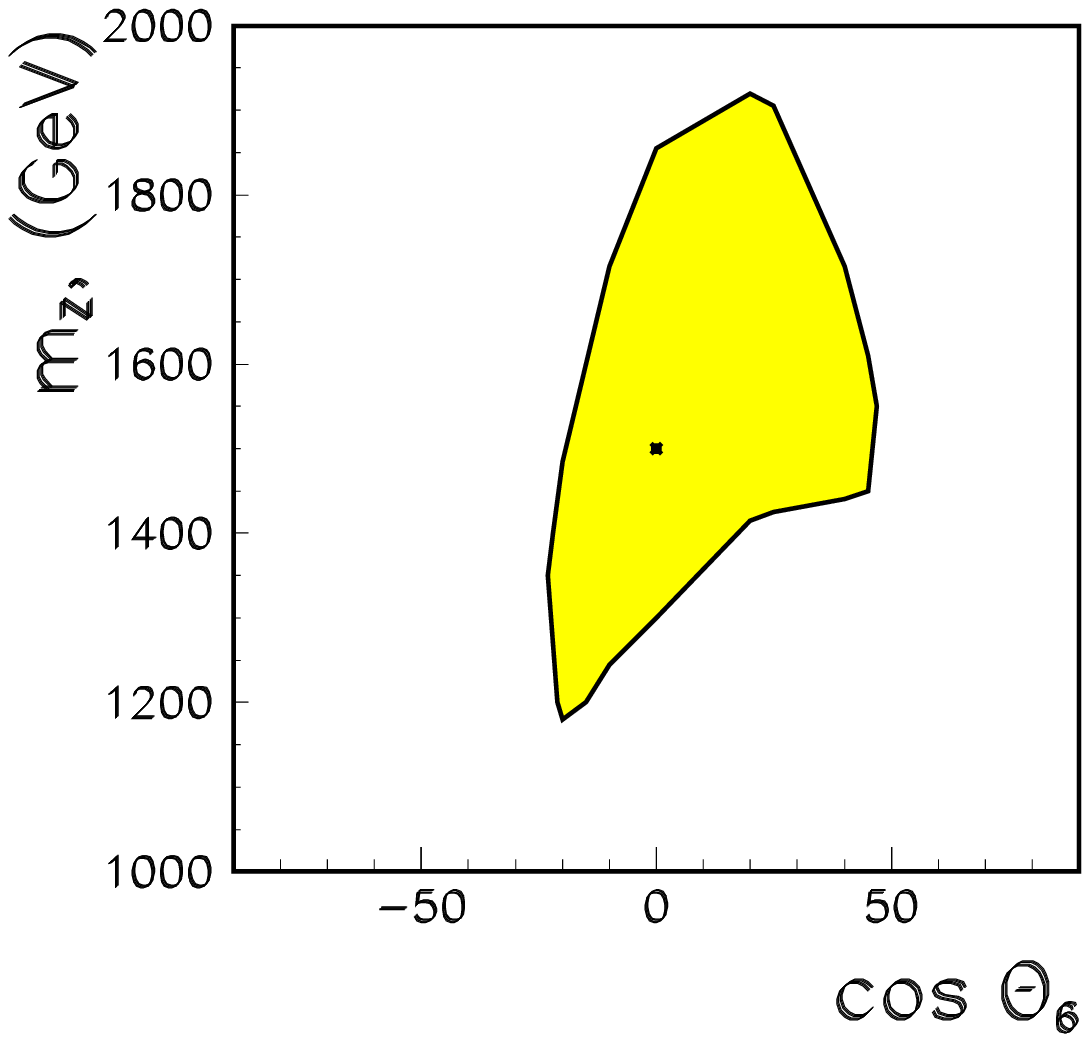}
 }
\end{tabular}
\end{flushleft}
\caption{The 95\%  CL
areas of $(m_{\Zp};~\Theta_6$) values for detecting the Z$'$ in the
$\chi$ model for
$m_{\Zp}=1$~TeV and $m_{\Zp}=1.5$~TeV;
$L_{int}=50$~fb$^{-1}$, $\sqrt{s}$=500 GeV.
}
\label{fig:2}
\end{figure}
\begin{figure} 
  \begin{center}
   \resizebox{!}{8cm}{%
   \includegraphics{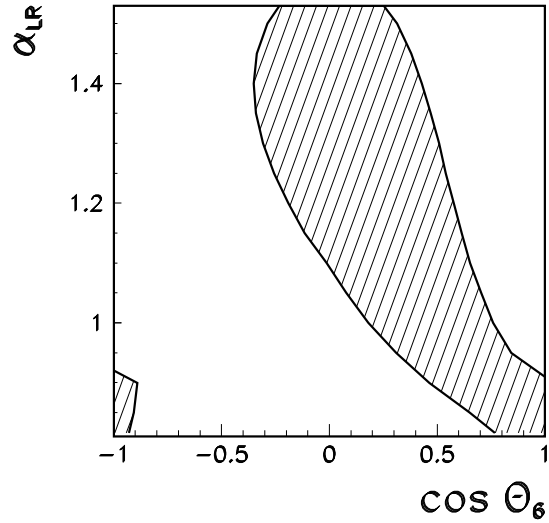}
}
 \end{center}
\caption{Confusion regions between E$_6$ and left-right models for
$m_{\Zp}=1.5$~TeV, $\sqrt{s}=500$~GeV, and $L_{int}$=20 fb$^{-1}$
based on $\sigma_l,~A_{FB}^l$ and $A_{LR}^l$.
}
\label{fig:3}
\end{figure}
If $m_{\Zp} \approx 2~\sqrt{s}$ the Z$'$ model {\it and} the Z$'$ mass
can be reproduced with a good precision. For $m_{\Zp} \approx 3~\sqrt{s}$
the determination of the Z$'$ model becomes difficult. Having in mind
that the $\chi$ model ($\Theta_6=0$) is identical with 
the left-right model for
$\alpha_{LR}=\sqrt{2/3}$, a confusion between Z$'$ models remains.
Figure \ref{fig:3} shows corresponding regions of confusion between
E$_6$ and left-right models assuming $m_{\Zp}=1.5$~TeV and
$L_{int}=20$~fb$^{-1}$.

\subsection{Z$'$ Couplings to Fermions}

If deviations from the Standard Model predictions will be found
the search for explanations will also include the  possibility that
a Z$'$ is the source of the disagreement.
But, are we able to analyze the Z$'$ without any knowledge about its
origin?

The measurements of 2-fermion final states
below the Z$'$ resonance
are sensitive to normalized Z$'$ couplings $a_f^N, v_f^N$ \cite{al},
\begin{eqnarray}
 a^N_f &= &a'_f\sqrt{s/(m_{Z'}^2-s)}, \nonumber \\
 v^N_f &= &v'_f\sqrt{s/(m_{Z'}^2-s)} \label{normcoup}.
\end{eqnarray}
From (\ref{normcoup}),
for a given Z$'$ mass  the couplings $a'_f$ and $v'_f$ can be
found and thus the Z$'$ model can be determined.
First,
let us assume that the Z$'$ is detected
at LHC and the Z$'$ mass is known.

\subsubsection{Z$'$ Couplings to Leptons}

Assuming lepton universality the situation is quite clear:
The  couplings of the Z$'$ to the initial and final states
are identical and can be determined with a good accuracy if the
Z$'$ is not too heavy compared to the collider energy.
\begin{figure} 
\begin{tabular}{ll}
\hspace{-1cm}

   \resizebox{!}{7cm}{%
   \includegraphics{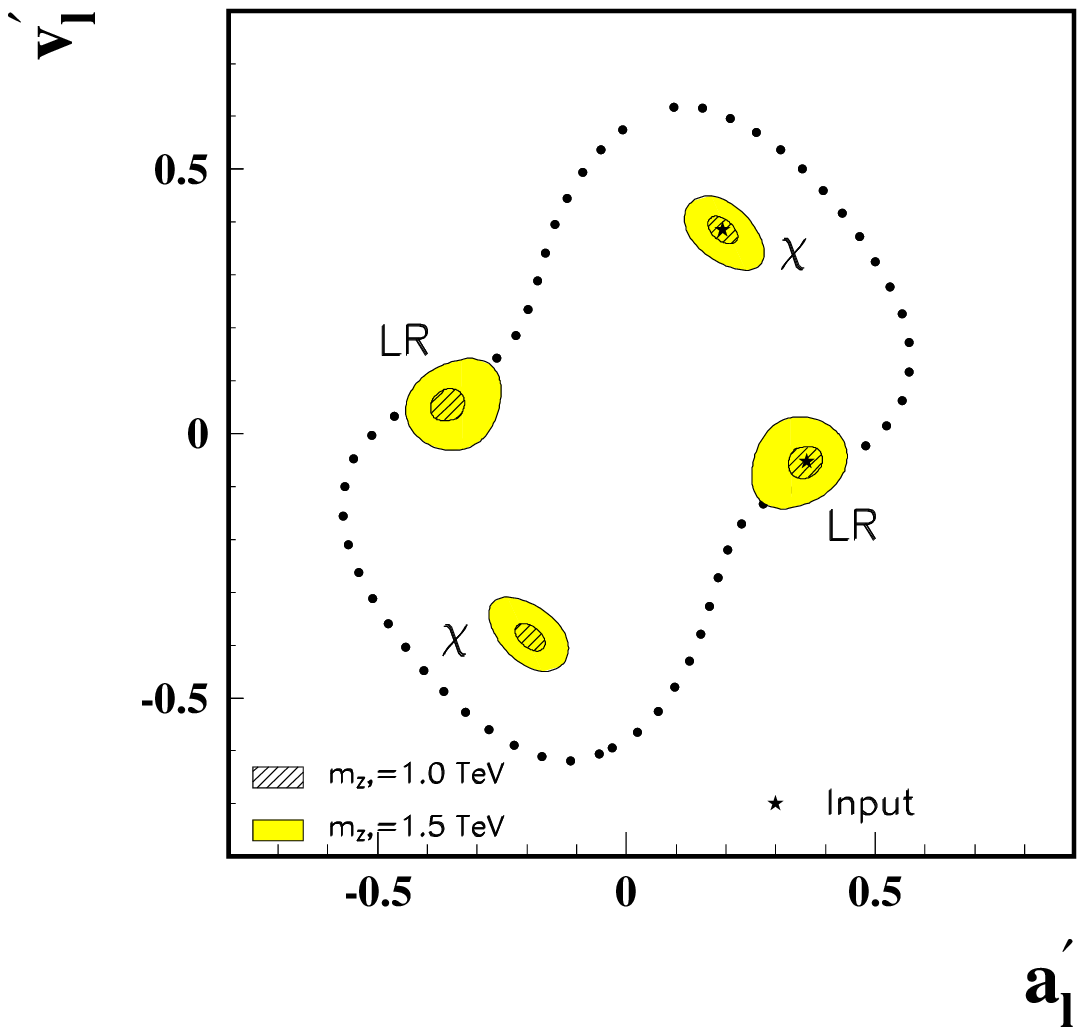}
}                     &
\hspace{-.51cm}

   \resizebox{!}{7cm}{%
   \includegraphics{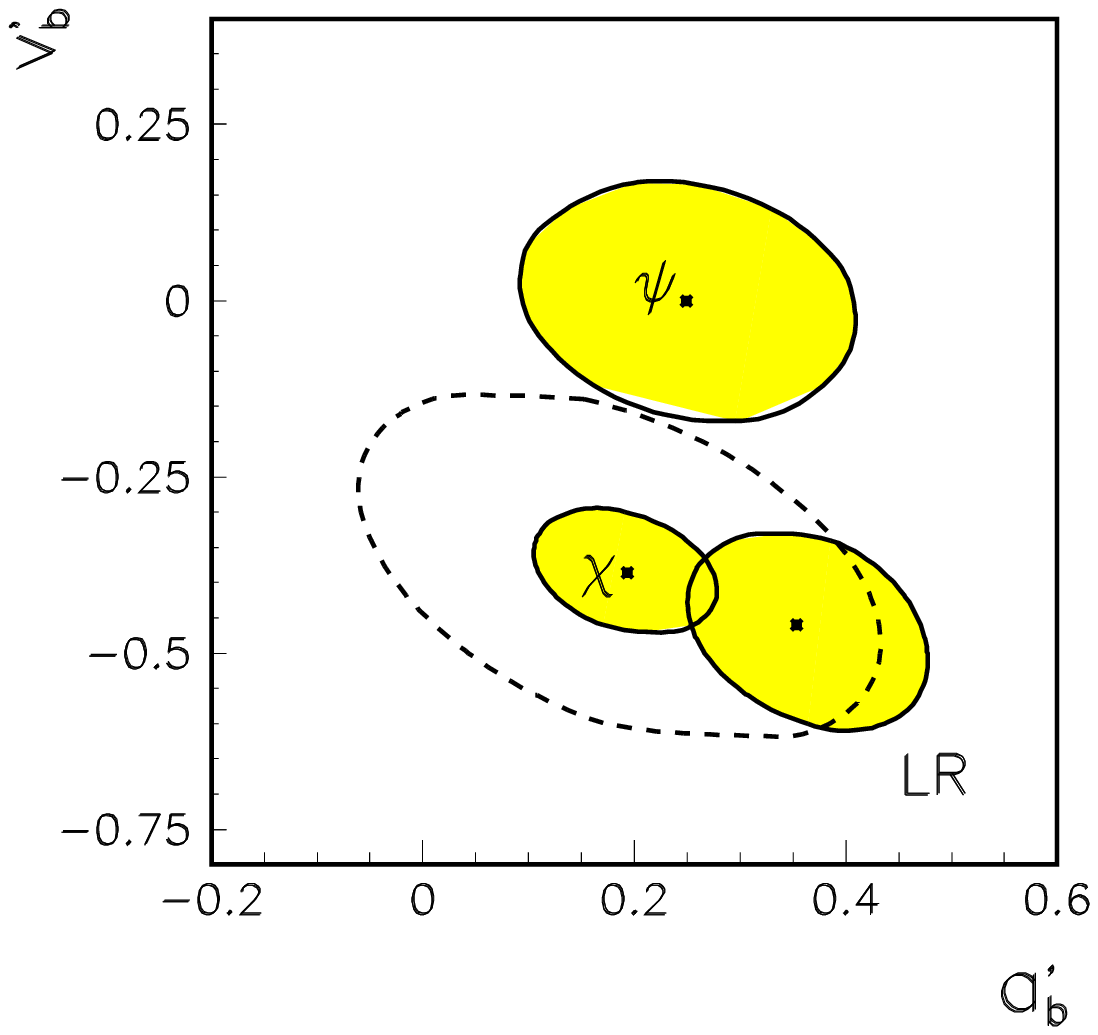}
}
\end{tabular}
\caption{(a) 95\% CL contours for $a'_l$ and $v'_l$. A Z$'$ is assumed
in the $\chi$ model or in the LR model
with a mass of  $m_{Z'}$=1 TeV (hatched area) and
$m_{Z'}$=1.5 TeV (shaded area).  The dotted line limits the
95\% CL bounds  on Z$' \mathrm{l} \bar{\mathrm{ l}}$ couplings 
if a Z$'$ with a mass $m_{Z'}$=3 TeV exists in the $\chi$ model.
${\cal L}=50$~fb$^{-1}$ and $\sqrt{s} = 500$~GeV.
(b) Discrimination between $\chi$, $\psi$ and LR model based on
95\% CL contours for $a'_b$ and $v'_b$ assuming $m_{Z'}$=1~TeV 
(hatched area).
For comparison a Z$'$, $m_{Z'}$=1.5~TeV, is considered in the $\chi$
model, too (dashed line).
Collider scenario (\ref{coll}) is taken into account.
}
\label{fig:4}
\end{figure}
The observables depend only on bilinear
products of $a'_f$ and
$v'_f$. Thus, a two-fold ambiguity in the signs of couplings remains.

Figure \ref{fig:4} shows the bounds on
Z$' \mathrm{l} \bar{\mathrm{l}}$  couplings
for different collider scenarios and various Z$'$ masses.
Evidently,
the sensitivity to the Z$'$ couplings is weakened
with decreasing luminosity and with increasing $m_{\Zp}$:
\begin{equation}
\frac{\Delta a'_1}{\Delta a'_2};~ \frac{\Delta v'_1}{\Delta v'_2}
\approx \displaystyle{ \left(
\frac{m_{Z'_1}^2-s}{m_{Z'_2}^2-s} \right) ^{1/2}},
\label{scal_lept}
\end{equation}
\begin{equation}
\frac{\Delta a'_1}{\Delta a'_2};~ \frac{\Delta v'_1}{\Delta v'_2} 
\approx   \left(
\frac{{\cal L}_2}{{\cal L}_1} \right)^{1/4}.
\label{eq_lumi}
\end{equation}
If $m_{Z'} \ge 5\cdot \sqrt{s}$ the Z$'$  influences
the observables only weakly and deviations from the Standard
Model predictions cannot be safely observed.
It is impossible to
exclude the point $(a'_l, v'_l) = (0, 0)$ in Figure \ref{fig:4}
with ~95\% CL, although
the existence of a Z$'$ ($\chi$ model) is assumed.
Even the indirect detection
of a Z$'$ is no longer possible.
Nevertheless,upper limits on Z$'$ couplings can be derived.

\subsubsection{Z$'$ Couplings to Quarks}

The determination of  Z$'$ couplings to quarks
depends on the knowledge of the couplings to electrons.
In particular, if the error range 
of $a'_e$
and $v'_e$ includes $a'_e=v'_e=0$, a simultaneous fit to leptonic
and quarkonic couplings will fail. In the following, we assume that
an analysis of leptonic observables leads to non-vanishing
Z$ \mathrm{l} \bar{\mathrm{l}}$ couplings.

The identification of quark flavors is more complicated than
lepton identification. 
Although very promising designs of a vertex detector for the LC
let us expect  efficiencies of more than 60\% in b--tagging
with  purities of at least 98\%  (see \cite{lc_vertex_design}),
the systematic error for the measurement of b-quark observables
could be about 1\% and dominate the statistical error.
The systematic
errors limit the accuracy of a $a'_q, v'_q$ determination
substantially and could fully remove improvements
due to a higher luminosity.
The promising model identification power
using the parameters,
\bq
P_V^l = \frac{v'_l}{a'_l},~~~~ P_L^q = \frac{v'_q+a'_q}{a'_l},~~~~
P_R^{u,d} = \frac{v'_{u,d}-a'_{u,d}}{a'_l+v'_l},  \label{par_cvetic}
\eq
as suggested by
\cite{delaguila-cvet} looses its charm  
if in addition to statistical errors realistic systematic errors
are taken into account (see Table 2 of \cite{alsr}).

\begin{figure} 
  \begin{center}
   \resizebox{!}{8cm}{%
   \includegraphics{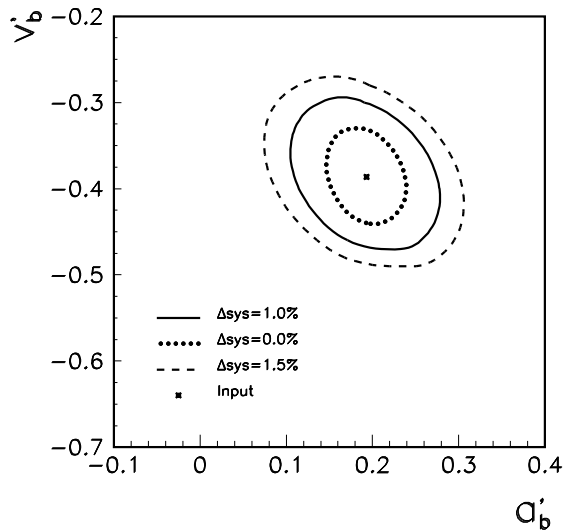}
}
 \end{center}
\caption{Influence of luminosity,  Z$'$ mass, and systematic error
 on contours of Z$' b \bar b$ couplings.
A Z$'$ in the $\chi$ model is assumed.}
\label{fig:5}
\end{figure}

In Figure \ref{fig:5} it is demonstrated for  the $\chi$ model
how  systematic errors
could complicate the determination of the
Z$' \mathrm{q} \bar{\mathrm{q}}$ couplings.
Otherwise,
comparing Figures \ref{fig:4} and \ref{fig:5} it is obvious
that the crucial quality for the determination of Z$'$ couplings is
the difference $m_{\Zp}-\sqrt{s}$.

\subsection{Z$'$ Couplings without Information about the Z$'$ mass}

\begin{figure} 
\begin{tabular}{ll}
\hspace{-1cm}

   \resizebox{!}{7cm}{%
   \includegraphics{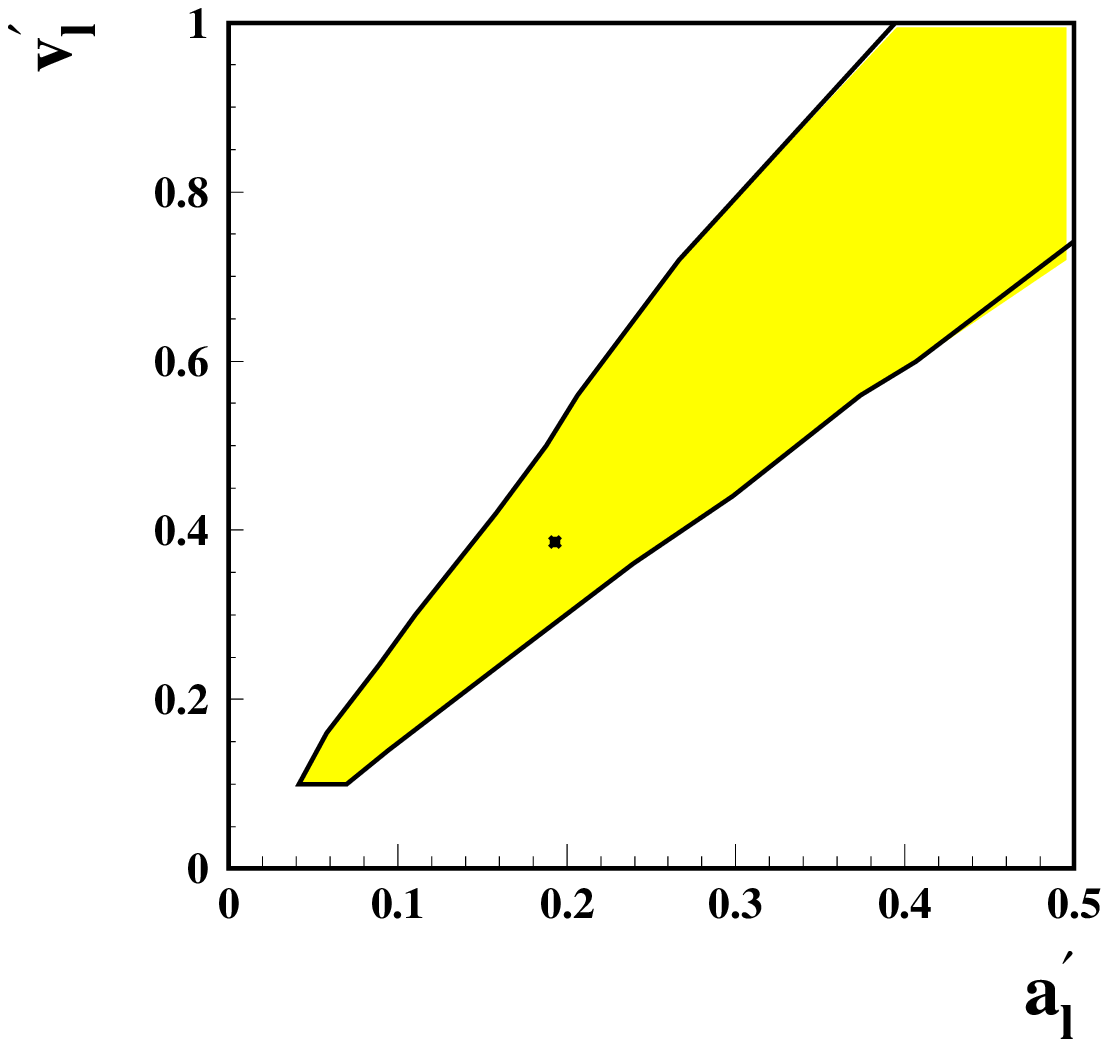}
}             &
\hspace{-.51cm}

   \resizebox{!}{7cm}{%
   \includegraphics{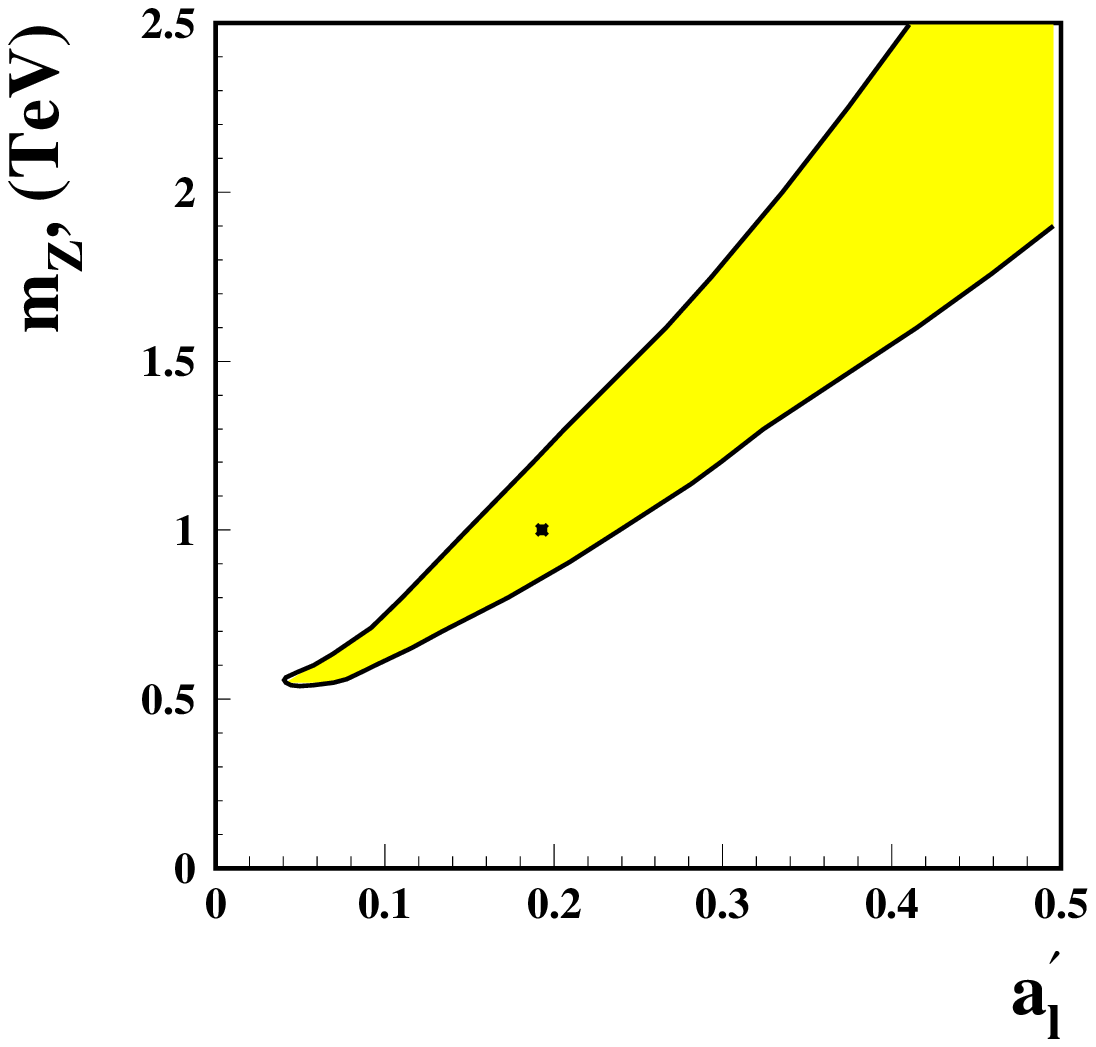}
}
\end{tabular}
\caption{95\% CL contours for $a'_l$, $v'_l$ and $m_{\Zp}$.
A Z$'$, $m_{\Zp}$=1~TeV,  assumed in the $\chi$ model ($\ast$)
 should be
reconstructed based on collider scenario (\ref{coll}).
Only the positive couplings are shown.
}
\label{fig:6}
\end{figure}

If the Z$'$ mass is unknown the determination of Z$'$ couplings
becomes difficult.
This is demostrated in Figure \ref{fig:6} assuming $m_{\Zp}$
= 1 TeV  in the $\chi$ model and studying leptonic observables only.
It is impossible to get
upper limits on Z$'$ mass {\it and} couplings simultaneously.
The mean axis of the ($a'_l,~v'_l$) contour in Figure \ref{fig:5}
corresponds to the above mentioned parameter $P_V^l = v'_l/a'_l$
and allows to some extent conclusions on the Z$'$ model.
Including $\mathrm{q} \bar{\mathrm{q}}$ final  states,
this method -- to extract information about the Z$'$ model using
the parameters $P_V,~P_R,~ P_L$ of Equ. (\ref{par_cvetic}) --
 fails due to  possible infinite Z$' \mathrm{l}
\bar{\mathrm{l}}$ couplings.

The boundaries of the ($a'_l, m_{\Zp}$) contour follow
approximately the relation
\bq
m_{\Zp}^{\pm}(s) = \sqrt{ \left[ \frac{a'^2_l}{(a^N_l\mp \Delta a^N_l)^2}
 + 1 \right] ~s}.
\eq
An additional
measurement at a higher energy, $s_2 > s_1$ resulting in
$ m_{\Zp}^{\pm}(s_2)$ can close the contours in Figure \ref{fig:6}
if
\bq
\left(\frac{s_1}{s_2}\right)^{1/2} = 
\frac{a_{l,1}^N - \Delta a_{l,1}^N}
     {a_{l,2}^N + \Delta a_{l,2}^N}
              \label{zp_crossing}
\eq
A scanning strategy considering (\ref{zp_crossing})
will allow the simultaneous determination of Z$'$ couplings and mass.

The success and failure of fitting Z$'$ couplings and mass
below a Z$'$ peak
is illustrated by Rizzo studying the model resolution
for various distributions
of luminosity on several energy points \cite{rizzo_coupl}.

\section{Conclusions}

If a Z$'$ boson with a mass  $m_{Z'} < 5 \sqrt{s}$
exists
observables measured at LC differ from their
Standard Model expectations.
The interpretation of these deviations within special Z$'$ models
gives the Z$'$ mass with a good accuracy or allows to determine
lower bounds on  the Z$'$ mass.
More interesting is a model-independent analysis.
With the determination of
Z$' \mathrm{f} \bar{\mathrm{f}}$ couplings  conclusions on the
Z$'$ model are possible.
If the Z$'$ mass is known and  $m_{Z'} < 3  \sqrt{s}$,
Z$'$ models can be separated well considering
lepton pair production only.
In case of q$\bar{\mathrm{q}}$ final states the accuracy of the
$a'_q,~v'_q$
coupling measurement is diminished by
the uncertainty of $a'_l,~v'_l$ and by
systematic errors which could reach the
magnitude of the statistical errors. A good model resolution is expected
for  $m_{Z'} < 2 \sqrt{s}$ for the considered collider scenario.
If the Z$'$ mass is unknown, only for energies $\sqrt{s}$ near the Z$'$
resonance a good indirect analysis of Z$'$ parameters is possible.
To analyze the Z$'$ below the Z$'$ production threshold
with a linear collider only, a special
scanning strategy is  essential.

\end{document}